\documentclass[11pt,a4paper]{article}
\usepackage{amsmath,amsfonts,amssymb,amsbsy}
\usepackage{mathrsfs,latexsym}
\usepackage{graphicx}
\usepackage{xcolor}
\usepackage{booktabs}
\usepackage{multirow}
\usepackage{multicol}
\usepackage{subfig}
\usepackage{placeins}
\usepackage{blindtext}
\usepackage{bm}
\usepackage{times}
\usepackage{float}
\usepackage{booktabs}
\usepackage{wrapfig}
\usepackage[utf8]{inputenc} 
\usepackage[T1]{fontenc} 
\usepackage{amsmath,amssymb}
\usepackage{alpha}
\usepackage{macros_alpha}

\begin{document}

\title{
\begin{flushright}
\small{
WUB/19-00\\
CP3-Origins-2019-002 DNRF90\\
MITP/19-010 \\}
\vskip 0.7cm
\end{flushright}
String breaking by light and strange quarks in QCD}

\author[label1]{John Bulava}
\author[label2,label3]{Ben H\"orz}
\author[label4]{Francesco Knechtli}
\author[label4,label5]{Vanessa Koch\corref{cor1}}
\ead{kochv@maths.tcd.ie}
\author[label6]{Graham Moir}
\author[label7]{Colin Morningstar}
\author[label5]{Mike Peardon}

\address[label1]{CP3-Origins, University of Southern Denmark, Campusvej 55, 5230 Odense M, Denmark}
\address[label2]{PRISMA Cluster of Excellence and Institute for Nuclear Physics, Johannes Gutenberg-Universit\"at, 55099 Mainz, Germany}
\address[label3]{Nuclear Science Division, Lawrence Berkeley National Laboratory, Berkeley, CA 94720, USA}
\address[label4]{Dept. of Physics, University of Wuppertal, Gaussstrasse 20, D-42119 Germany}
\address[label5]{School of Mathematics, Trinity College Dublin, Dublin 2, Ireland}
\address[label6]{Dept. of Mathematics, Hurstpierpoint College, College Lane, Hassocks, West Sussex, BN6 9JS, United Kingdom}
\address[label7]{Dept. of Physics, Carnegie Mellon University, Pittsburgh, PA 15213, USA}

\begin{abstract}
The energy spectrum of a system containing a static quark anti-quark pair
is computed for a wide range of source separations using lattice QCD with 
$N_\mathrm{f}=2+1$ dynamical flavours. By employing a variational 
method with a basis including operators resembling both the gluon string and 
systems of two separated static mesons, the first three energy levels are 
determined up to and beyond the distance where it is energetically favourable 
for the vacuum to screen the static sources through light- or strange-quark 
pair creation, enabling both these screening phenomena to be observed.
The separation dependence of the energy spectrum is reliably parameterised 
over this saturation region with a simple model which can be used as input for 
subsequent investigations of quarkonia above threshold and heavy-light and 
heavy-strange coupled-channel meson scattering. 
\end{abstract}

\maketitle

\section{Introduction}
\label{sec:intro}

QCD is believed to be responsible for confinement, the experimental observation
that quarks are never seen as asymptotic states \cite{Wilson:1974sk}. A full understanding of why
QCD confines remains elusive. The simplest theoretical probe of the
phenomenon is provided by the potential energy $V(r)$ of a system made of a static
quark and anti-quark pair immersed in the QCD vacuum in a colourless
combination. This energy depends on the distance between the sources, and in the
Yang-Mills theory of gluons alone it grows linearly at asymptotically large separations.
The rate of increase is the well-known string tension, $\sigma$.  If
the static sources interact with the full QCD vacuum including light-quark
dynamics, pair-creation of light quarks means the system can also resemble two
separately-colourless static-light mesons allowing the potential energy to 
saturate at large separations. This phenomenon, induced by light-quark pair 
creation, is termed \lq string breaking\rq.

Arising from purely non-perturbative effects, the static potential requires a 
robust method for direct determination from the QCD Lagrangian. Lattice QCD 
provides such a framework, but studying string breaking on the lattice is a 
technical and numerical challenge. The simplest approach would be to evaluate 
the expectation value of large Wilson loops in the QCD vacuum and observe 
deviation from an area law. However, Monte Carlo determinations of large loops 
suffer 
from poor signal-to-noise properties. Also, viewed in terms of eigenstates of 
the Lattice QCD Hamiltonian, the creation operator comprising the Wilson line forming one 
edge of the rectangular Wilson loop has very small overlap onto the 
two-meson system dominating the ground state at large separations. As a consequence, the saturation 
effect proves impossible to resolve from Monte Carlo studies of Wilson 
loops alone. Instead, a mixing analysis including two-meson \lq broken\rq ~string 
states is needed \cite{Drummond:1998ar,Philipsen:1998de,Knechtli:1998gf,Knechtli:2000df,bali}.

In this work, by building a suitably diverse basis of creation operators, 
mixing between the state made by a gluonic flux tube and the broken string 
state resembling two static-light or static-strange mesons is investigated 
fully in the $N_\mathrm{f}=2+1$ theory on the lattice for the first time. This enables 
us to 
compute reliably the energies of the lowest three Hamiltonian eigenstates up to 
separations where string breaking saturation occurs. A simple 
parameterisation of the resulting spectrum in the breaking region is 
given, which should provide invaluable first-principles input into models of 
coupled-channel scattering of heavy-light mesons and the decays of quarkonia 
near threshold. 

\section{Methodology}

\label{sec:methods}
We compute the potential energy of a system containing a heavy quark 
$Q$ at spatial position $\mathbf{x}$ and a heavy anti-quark 
$\bar{Q}$ at $\mathbf{y}$ in the static approximation. 
In this limit, the quarks remain separated by 
$\mathbf{r}=\mathbf{y}-\mathbf{x}$, where $\mathbf{x}$ and $\mathbf{y}$ 
are conserved quantum numbers. 
To determine the energies of the ground state as well as the first- and 
second-excited state arising from mixing in QCD, a variational technique 
is employed. The input for this mixing calculation is a matrix of temporal 
correlation functions between the interpolator 
for a Wilson line $\mathcal{O}_W$, the two-static-light 
$\mathcal{O}_{B\bar{B}}$ and the two-static-strange meson states
$\mathcal{O}_{B_s\bar{B}_s}$. It is essential that the interpolators transform 
irreducibly under the appropriate symmetry group. 

\subsection{Interpolators}

A suitable interpolator $\mathcal{O}_W$ creating a gluon string connecting 
sites $\mathbf{x}$ and $\mathbf{y}$ at time $t$ is given by
\begin{align}
\mathcal{O}_W (\mathbf{y},\mathbf{x},t)\ \ &= \ \ \bar{Q}(\mathbf{y},t) \frac{\boldsymbol{\gamma}\cdot\mathbf{r}}{r}   \mathcal{W}(\mathbf{y},\mathbf{x},t) Q(\mathbf{x},t) ,
\end{align}  
where $\boldsymbol{\gamma}$ is the three-vector of spatial Dirac matrices and 
the Wilson line $\mathcal{W}(\mathbf{y},\mathbf{x},t)$ is a product of 
spatial lattice links with time argument $t$ connecting 
$\mathbf{x}$ and $\mathbf{y}$. 
Wilson lines are constructed using a variant of the Bresenham algorithm 
\cite{Bresenham} to approximate the shortest connection between 
$\mathbf{x}$ and $\mathbf{y}$, cf. \cite{Bolder:2000un}. 
The heavy-quark spins are coupled symmetrically via 
$\boldsymbol{\gamma}\cdot\mathbf{r}/r$, which has a zero component of angular
momentum projected along $\mathbf{r}$. In the static limit, both the
symmetric and antisymmetric combinations give an energy level in the 
$\Sigma_g^+$ irreducible representation of the rotation group around 
$\mathbf{r}$ after the heavy spins are decoupled. Decoupling the heavy spins 
in the two-static-meson system similarly yields a composite operator in the 
$\Sigma_g^+$ irrep. Details of this construction can be found in 
Ref.~\cite{bali}. 
With light-quark flavours $q^i, i=\{u,d,s\}$, suitable interpolators for a 
two-static-light- or two-static-strange-meson state are given by 
\begin{align} \label{e:two-meson-interp}
 \mathcal{O}_{B\bar{B}} (\mathbf{x},\mathbf{y},t)\ \ &=\frac{1}{\sqrt{2}}
  \sum_{i=u,d}&\bar{Q}(\mathbf{y},t) \gamma_5 q^i(\mathbf{y},t) \ \bar{q}^i(\mathbf{x},t) \gamma_4\gamma_5 Q(\mathbf{x},t),\nonumber \\
 \mathcal{O}_{B_s\bar{B}_s} (\mathbf{x},\mathbf{y},t)\ \ &=&\bar{Q}(\mathbf{y},t) \gamma_5 q^s(\mathbf{y},t) \ \bar{q}^s(\mathbf{x},t) \gamma_4\gamma_5 Q(\mathbf{x},t).
\end{align}
For the two-static-light-meson state, the sum 
projects onto the isospin-zero channel.  
Notice that our interpolators Eq.~(\ref{e:two-meson-interp}) contain a $\gamma_4$
because for the inversion of the Dirac operator we use the convention of Ref. \cite{morningstar}.

\subsection{Correlation matrix}
For the $N_\mathrm{f}=2+1$ theory, a $3\times 3$ matrix can be constructed from the 
pair-wise correlations of the two isospin-zero meson-pair creation and 
annihilation operators in combination with the string interpolation operator 
\begin{align}\label{sbmat}
C(\mathbf{r},t)=& 
  \left(\begin{array}{ccc}
    \langle\mathcal{O}_{W}(t)            \overline{\mathcal{O}}_{W}(0)\rangle
  & \langle\mathcal{O}_{B\bar{B}}(t)     \overline{\mathcal{O}}_{W}(0)\rangle
  & \langle\mathcal{O}_{B_s\bar{B}_s}(t) \overline{\mathcal{O}}_{W}(0)\rangle 
\\ 
    \langle\mathcal{O}_{W}(t)            \overline{\mathcal{O}}_{B\bar{B}}(0)\rangle
  & \langle\mathcal{O}_{B\bar{B}}(t)     \overline{\mathcal{O}}_{B\bar{B}}(0)\rangle
  & \langle\mathcal{O}_{B_s\bar{B}_s}(t) \overline{\mathcal{O}}_{B\bar{B}}(0)\rangle
\\
    \langle\mathcal{O}_{W}(t)            \overline{\mathcal{O}}_{B_s\bar{B}_s}(0)\rangle 
  & \langle\mathcal{O}_{B\bar{B}}(t)     \overline{\mathcal{O}}_{B_s\bar{B}_s}(0)\rangle 
  & \langle\mathcal{O}_{B_s\bar{B}_s}(t) \overline{\mathcal{O}}_{B_s\bar{B}_s}(0)\rangle 
\end{array} \right).
\end{align}
When mixing occurs, the above basis states are not energy eigenstates anymore and 
off-diagonal elements of the correlation matrix are non-vanishing.
Following the method presented in \cite{donnellan}, 
all gauge-links are smeared using HYP2 parameters \cite{hyp2,dellamorte} 
$\alpha_1 = 1.0, \alpha_2 = 1.0, \alpha_3 = 0.5$, where
the smearing of the temporal links amounts to a modification of the Eichten-Hill static 
action and propagator \cite{eichtenhill} which reduces the divergent mass 
renormalisation and improves the signal-to-noise ratio at large Euclidean 
times \cite{dellamorte}. As a second step, we construct a variational basis for the string state using 15 and 20 levels 
of HYP-smeared spatial links with parameters: $\alpha_2 = 0.6, \alpha_3 = 0.3$, extending 
Eq.~(\ref{sbmat}) to a $4 \times 4$ matrix.

Some correlation functions in $C(\mathbf{r},t)$ include multiple 
light-quark field insertions. The light-quark fields must be integrated 
analytically prior to Monte Carlo evaluation of $C(\mathbf{r},t)$ and the 
resulting Wick contractions involve numerically challenging disconnected 
contributions. 
In order to calculate these contributions, and to reduce statistical variance 
by exploiting translational invariance of the lattice, propagators between 
all space-time points are needed. For this, we employ the stochastic LapH 
method \cite{morningstar}.  Based on distillation \cite{distillation}, the 
method facilitates all-to-all quark propagation in a low-dimensional subspace 
spanned by $N_v$ low-lying eigenmodes of the three-dimensional gauge-covariant 
Laplace operator, which is constructed using stout-smeared gauge links 
\cite{stout} with parameters $\rho=0.1,n_{\rho}=36$. This projection onto the 
so-called LapH subspace amounts to a form of quark smearing, where $N_v$ 
increases in proportion to the spatial volume for a fixed physical smearing 
radius.  Introducing a stochastic estimator in the LapH subspace in combination 
with dilution \cite{dilution2,dilution} 
helps to reduce the rise in computational costs as the volume increases. It 
was shown \cite{morningstar} that for 
certain dilution schemes, the quality of the stochastic estimator remains 
approximately constant for increasing volume, while maintaining a fixed number 
of dilution projectors. The triplet $b=(T,S,L)$ specifies a dilution scheme with time $T$, spin $S$ and 
LapH eigenvector projector index $L$, where $F$ indicates full dilution and $In$ the interlacing of dilution projectors in index $n$. 
The dilution scheme and other stochastic-LapH parameters used are given in 
table \ref{table:param}.
\begin{table}[h]
  \centering
  \begin{tabular}{c c c c c c c}
    \toprule
     & & & \multicolumn{2}{c}{light} & \multicolumn{2}{c}{strange} \\
     \cmidrule(lr){4-5}
     \cmidrule(lr){6-7}
     Type & Dilution scheme & Source times, $t_s/a$ & $N_r$ & $n_\mathrm{inv}$ & $N_r$ & $n_\mathrm{inv}$ \\
     \midrule
     fixed & (TF,SF,LI8) & \{32, 52\} & 5 & 320 & 2 & 128 \\
     relative & (TI8,SF,LI8) & \{32, 33, \dots, 95\} & 2 & 512 & 1 & 256 \\
     \bottomrule
  \end{tabular}
\caption{
The parameters of the quark-line estimation method employed in this work. $N_v=192$ eigenvectors were used to form the distillation operator. 
Quark lines starting and terminating at the same time slice $t_s=t_{\!f}$ are referred to as \lq relative\rq, while quark lines with $t_s\neq t_{\!f}$ are called \lq 
fixed\rq.
For definition of the LapH subspace and specification
of dilution schemes, see \cite{morningstar}. For these schemes, a
total of $32 \cdot 5 \cdot 2 + 32 \cdot 2 \cdot 8 = 832$ (light) and 
$32 \cdot 2 \cdot 2 + 32 \cdot 8 = 384$ (strange) solutions of the 
Dirac equation are required per gauge configuration, which however can be
reused for other spectroscopy projects, see e.g. \cite{andersen}.}\label{table:param}
\end{table}

\subsection{Variational analysis}\label{sec:gevp}

A correlation matrix is evaluated for a set of static-source separations
$\mathbf{r}$ and time separations $t$. The energy spectrum of the system for 
each spatial separation can be extracted by solving a 
generalised eigenvalue problem (GEVP) for each $\mathbf{r}$ 
\cite{gevp,luscher_gevp}, 
\begin{equation}\label{fullgevp}
 C(t)\, v_n(t,t_0) = \lambda_n(t,t_0)\, C(t_0)\,v_n(t,t_0) \,, 
  \quad n=1,\ldots,N\,,\quad t>t_0 ,
\end{equation}
where $\lambda_n$ and $v_n$ are the eigenvalues and eigenvectors respectively. 
After solving the GEVP, energies are extracted using a 
correlated-$\chi^2$ minimisation to a two-parameter, single-exponential fit 
ansatz. 

To simplify the analysis, the GEVP is first solved for a fixed pair of time 
separations, $(t_0,t_d)$ 
where $t_0$ is a reference separation and $t_d$ is the diagonalisation time
\cite{bulava_phaseshifts}. 
The correlation matrix of the resulting interpolators from this optimisation 
is then defined as 
\begin{equation}
 \hat{C}_{ij}(t)= \big( v_i(t_0,t_d), C(t)v_j(t_0,t_d)\big),
  \label{eq:cij}
\end{equation}
where the parentheses denote an inner product over the original operator 
basis. 
A potential source of systematic error is introduced, as off-diagonal elements 
of $\hat C_{ij}$ are not exactly zero. We control this by assessing the stability of the GEVP 
against varying the operator basis and using different pairs $(t_0,t_d)$.
Comparing the results for some sample distances to the GEVP as given in Eq.~(\ref{fullgevp}) shows 
agreement between both methods.

We are interested in the difference between the energies $V_n(r)$ of the ground $(n=0)$, 
first $(n=1)$ and second excited state $(n=2)$
and twice the energy of the static-light meson $2E_B$, which can be directly 
extracted from fits to the ratio \begin{align}\label{ratio}
 R_n(t)= \frac{\hat{C}_{nn}(t)}{C_B^2(t)} ,
\end{align}
of the diagonal elements of the rotated correlation matrix of Eq.~(\ref{eq:cij}) and
the correlation function of a single static-light meson squared $C_B^2(t)$.
The energy difference is extracted using a single-exponential fit. The fitted energies typically vary little as 
diagonalisation times $(t_0,t_d)$ or operator basis are varied.  
 
\section{Numerical results}
\label{sec:res}
Monte Carlo samples are evaluated on a subset of evenly-spaced configurations 
of the N200 CLS (Coordinated Lattice Simulations) ensemble with $N_\mathrm{f} = 2 + 1$ flavours of 
non-perturbatively  $O(a)$-improved Wilson fermions. The lattice size is 
$N_t \times N_s^3 = 128 \times 48^3$ with an estimated lattice spacing of 
0.064 fm and pion and kaon masses of $m_{\pi}=$ 280 MeV and $m_{K}=$ 460 MeV 
respectively \cite{cls}. Open temporal boundary conditions are imposed on the fields. 
The imprint on observables is expected to fall
exponentially with distance from the boundary \cite{openboundary} so 
measurements are made on the central half of the lattice only.

Previous studies found the signal quality to be limited by the Wilson-loop correlation functions ~\cite{bali},
which a preliminary analysis on a subset of our data corroborated. We mitigate this issue by measuring 
Wilson loops on 1664 configurations, while diagrams containing light or strange propagators are
evaluated on a subset of 104 samples.  The Wilson loops are then averaged into 104 
bins containing 16 configurations each, with the centre of the bin aligned 
with one entry in the 104-configuration subset. 
\begin{figure}[h]
\includegraphics[width=0.9\linewidth]{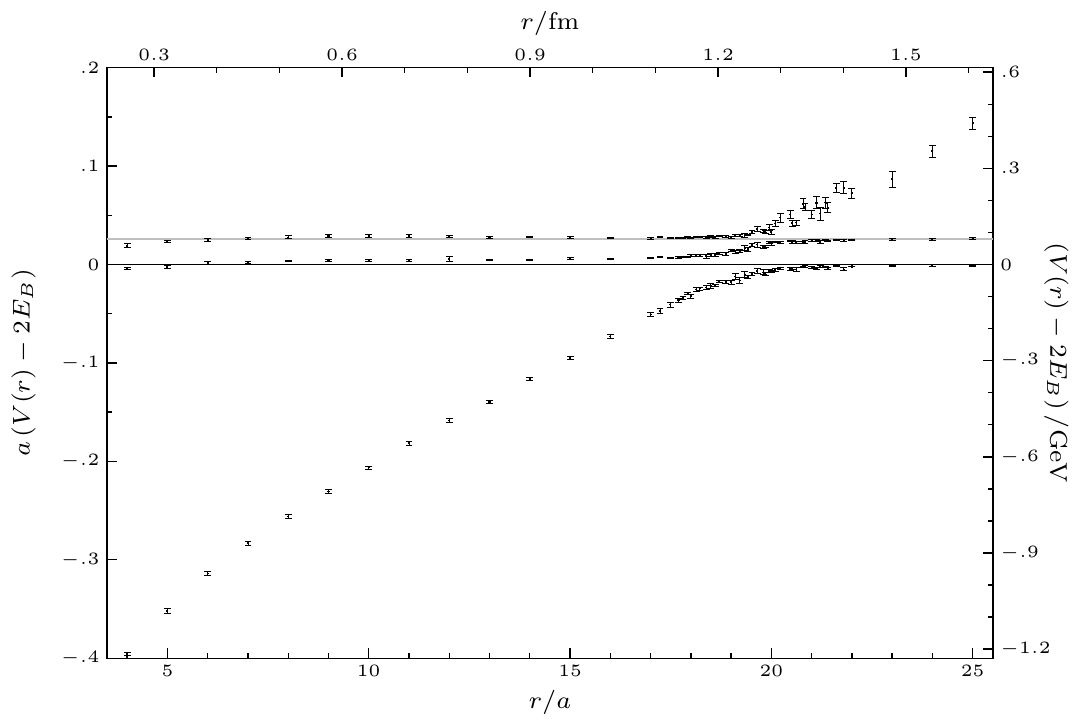}
\caption{Static potential determined using the full mixing matrix; the three lowest lying energy levels $V_n(r)$, $n=0,1,2$ are shown. The grey line corresponds to twice the static-strange meson mass, its error is too small to be visible. The black line corresponds to twice the static-light meson mass; the error is automatically taken into account by using the ratio given in Eq.~(\ref{ratio}). For all distances, the fixed GEVP with $t_0/a=5,t_d/a=10$ is used.}
\label{figure:full}
\end{figure}
For each separation $\mathbf{r}=(r_1,r_2,r_3)$, we exploit cubic symmetry and 
average over spatial rotations to increase our statistics.  As expected, no 
dependence on the direction is observed in our data. 
The analysis uses a Jupyter notebook\footnote{https://github.com/ebatz/jupan} 
adapted from Ref.~\cite{andersen}.
Statistical uncertainties are estimated using 800 bootstrap resamplings
\cite{bootstrap1,bootstrap2} and the uncertainty 
quoted is given by $1\sigma$  bootstrap errors. The covariance matrix entering the 
single exponential fits to Eq.~(\ref{ratio}) is estimated once and 
kept constant on every bootstrap sample.

In Fig.~\ref{figure:full}, the potential energy relative to $2E_B$ is 
shown; this subtraction removes the divergent mass renormalisation of a static 
quark.  
The grey line corresponds to twice the static-strange mass, with an energy difference 
$2E_{B_s}-2E_B=0.028(5)a^{-1}=85(16)$MeV. As expected due to the choice of light and strange quark masses, 
the difference is smaller than the physical energy difference between $B$ and $B_s$ mesons.

The avoided level crossing between the ground-
and first-excited states is clearly visible and the expected second avoided 
crossing due to the formation of two static-strange mesons is also evident for 
the first time. As the distance over which this 
phenomenon occurs is small, this effect cannot be resolved using only 
on-axis separations \cite{Koch:2018puh}.  For distances beyond the string breaking 
scale, the ground state tends rapidly towards the mass of two non-interacting 
static-light mesons. 

The matrix of correlation functions Eq.~(\ref{sbmat}) we use contains three very different 
operators, which should have strong overlap onto the three lowest physical energy 
eigenstates. Following Ref.~\cite{gevp}, we expect problems with determining energies 
from the GEVP using a finite basis will arise when higher states for which no good operator 
appears in the basis are close in energy. We can estimate where the next energy 
levels should be around the breaking region and they are all higher by a scale 
of about $500$MeV, substantially larger than the gaps observed.

The string breaking region is reproduced in more 
detail in Fig.~\ref{figure:model}. Both avoided crossings are visible and the energy gap between the ground state and 
first level is larger than the gap between first and second levels. 
Qualitatively, the first mixing region appears to be broader, but it is 
not possible to determine the difference between the first string breaking 
distance $r_c$ and the second string breaking distance $r_{c_s}$ by eye. 
The quantification of string breaking involving three levels is more complex in 
comparison to the two-level situation. For the $N_\text{f}=2$ vacuum, the 
string breaking distance $r_c$ can be defined by the minimum of the energy gap $\Delta E$ \cite{bali}. 
When the strange quark is included, an alternative definition of the two string breaking 
distances $r_{c}$ and $r_{c_s}$ is needed as there is not necessarily a minimum
energy gap.

\section{A model of the string breaking spectrum}

\label{sec:model}
\begin{figure}[tbp]
\begin{center}
\includegraphics[width=0.8\linewidth]{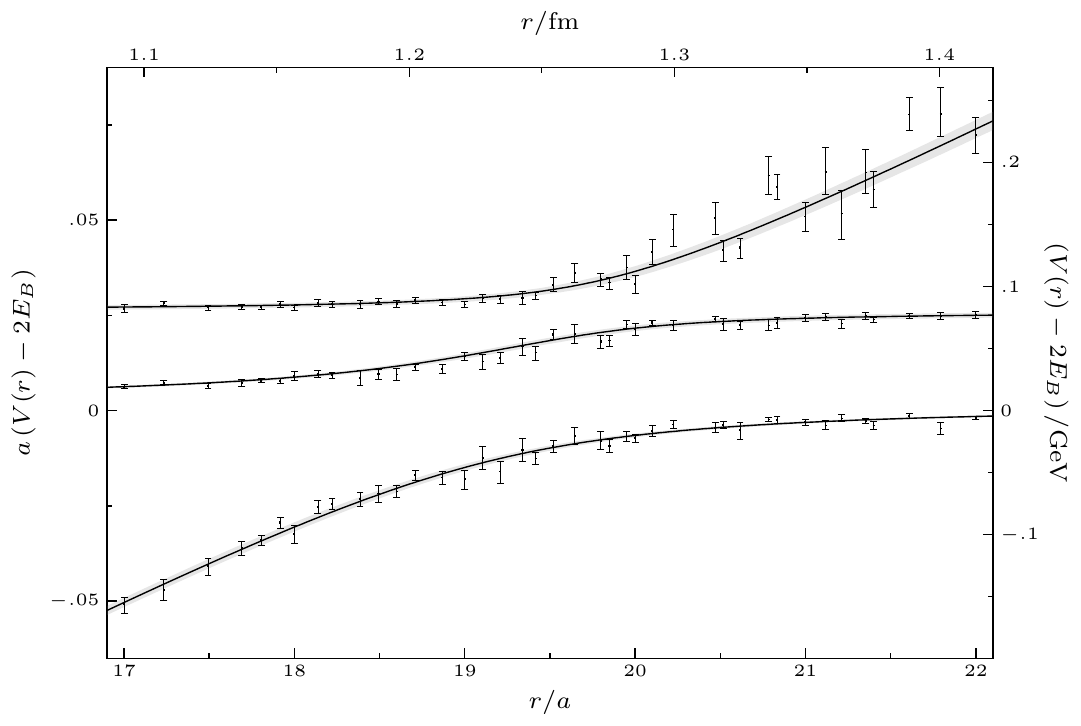}
\caption{Six-parameter fit to the string breaking data of Fig.~\ref{figure:full} over the fit range $r/a=[11,25]$. The error band indicates $1\sigma$ bootstrap errors.}\label{figure:model}
\end{center}
\end{figure}

We describe the potential-energy spectrum in the breaking region using a 
simple Hamiltonian that extends the model for $N_\mathrm{f}=2$ given in 
\cite{bookfrancesco}.
Consider a three-state system with Hamiltonian:
\begin{eqnarray}\label{e:model}
H(r)\quad=&&\left(\begin{array}{rcl}\hat{V}(r)& g_1 & g_2\\ g_1 & \hat E_1 & 0 \\ g_2 & 0 &\hat E_2\end{array}\right).\\ \nonumber
\end{eqnarray}
The diagonal elements are a function $\hat{V}(r)$ describing the unbroken 
string and $\hat E_1$, $\hat E_2$, the energies of a noninteracting pair of 
static-light and static-strange mesons, respectively. As with $\hat V$, these
energies are measured relative to $2E_B$. $g_1$ and $g_2$ are two coupling 
constants describing the strength of the mixing between the gluon flux and
the separated colour-screened static sources. The off-diagonal term that 
would mix the two-static-light-meson state with the two-static-strange-meson
state is set to zero. There is no constraint on this mixing in the energy 
spectrum alone as a basis rotation shows any non-zero value of this parameter 
yields an equivalent spectrum to the Hamiltonian of Eq.~(\ref{e:model}). 
Moreover, setting this value to zero ensures the diagonal elements of $H$ 
correspond to the asymptotic energy eigenvalues up to corrections at 
${\cal O}(r^{-1})$ in the limit $r\rightarrow\infty$.

A suitable choice for the function representing the string state is the Cornell potential \cite{cornell}. Since 
we are modelling the string breaking region and not the potential at 
small distances, only the linear part of the Cornell potential
\begin{equation}\label{cornell}
 \hat{V}(r)= \hat V_0 +\sigma r ,
\end{equation}
with string tension $\sigma$, is constrained by our data. Notice that we
assume that the model parameters $\hat E_1$, $\hat E_2$, $g_1$ and $g_2$
in Eq.~(\ref{e:model}) are
independent of the distance. We will see that this simplest possible choice
models our data very well in the region of string breaking.

The eigenstates of the Hamiltonian are mixtures of the unbroken string and the 
two static-light- and two static-strange-meson states while the eigenvalues of 
$H$ correspond to the three extracted energy levels. After diagonalising $H$, we 
perform an uncorrelated six-parameter fit to the spectrum, whose result is shown in Fig.~\ref{figure:model}.
We find for our fit parameters:
\begin{align}
 a\hat E_1 =&   0.0019(2),    \hspace{1cm} a\hat E_2 =  0.0262(6),   \\ \nonumber
 ag_1 =&   0.0154(4),    \hspace{1cm} ag_2 =   0.0080(5),     \\ 
 a^2\sigma =& 0.0229(3), \hspace{1cm} a\hat V_0 =  -0.434(5) .     \nonumber
 \end{align}

The model in Eq.~(\ref{e:model}) assumes a three state system. While it is possible that the physical eigenstates receive contributions from higher lying states Fig.~\ref{figure:model} shows that our data is well-described by the fit parameters.

\begin{figure}[tbp]
\begin{center}
\includegraphics[width=0.8\linewidth]{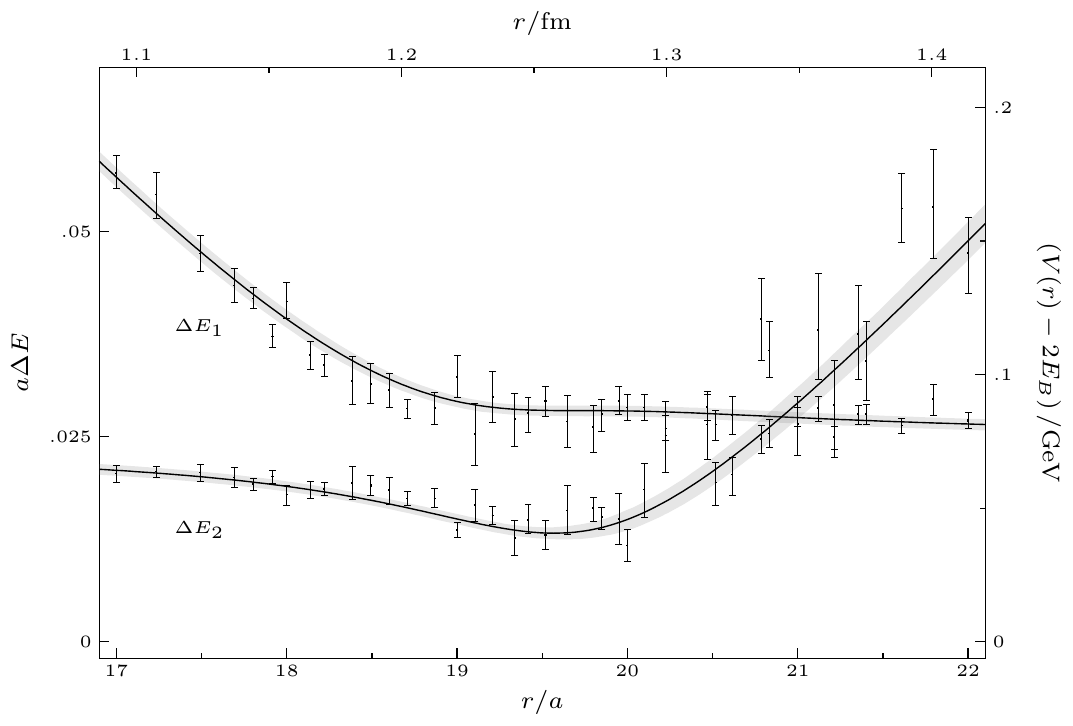}
\caption{Energy gap between ground and first-excited state $(\Delta E_1(r))$, as well as first-excited and second-excited state $(\Delta E_2(r))$ from the model fit as well as from our data. The error band and error bars indicate 
$1\sigma$ bootstrap errors. }\label{figure:gap}
 \end{center}
\end{figure}

We now turn to a quantitative definition of the string breaking distance in the $N_\mathrm{f}=2+1$ case. To investigate the 
dependence of the string breaking distance on the sea quark masses, a robust 
definition is needed. By using the asymptotic states of our model, which for large distances $r$ are given by 
the diagonal entries of our Hamiltonian, we extract two distinct string breaking distances corresponding to the light and strange mixing phenomenon. The string breaking distance, $r_c$
associated with the formation of two static-light quarks is given by the 
crossing distance where $\hat{V}(r_c)=\hat E_1$ and 
a corresponding definition can be employed to define $r_{c_s}$ using 
$\hat{V}(r_{c_s})=\hat E_2$. Our calculation yields 
\begin{align}
r_c = 19.053(82) a = 1.224(15) \text{fm} \ , \ \
r_{c_s} = 20.114(87)a  = 1.293(16) \text{fm}.
\end{align}
The quoted errors for the physical units take into account the uncertainty of $a=0.06426(76)$fm \cite{cls_scalesetting}. 

Fig.~\ref{figure:gap} shows that the energy gap $\Delta E_1(r)=V_1(r)-V_0(r)$ between the ground and first-excited state does not exhibit a minimum, thus making it impossible to use the minimal gap distance to define $r_c$. It can also be observed that in the string breaking region, $\Delta E_1(r)$ is twice as large as $\Delta E_2(r)=V_2(r)-V_1(r)$ , the energy gap between the first-excited and second-excited state.

Only one previous study \cite{bali} of string breaking for $N_\mathrm{f}=2$ QCD on the lattice exists with $m_{\pi}= 640$MeV. 
However, the differing definitions and quark content of the vacua mean it is not possible to make a statement on quark mass dependence. 
Using the position of the minimal energy gap, they find $\hat{r}_{c} \approx  1.248(13) \text{fm}$. Even though in this study the pion mass was relatively heavy, $\hat{r}_c$ is of the same order of magnitude as our result and falls between our values for $r_c$ and $r_{c_s}$. 

\subsection{Phenomenology}
To make a first, simple comparison between our data and the experimentally 
observed quarkonium spectrum, we use the ground-state energy computed from 
our model Eq.~(\ref{e:model}) as input potential in the Schr{\"o}dinger 
equation and extract the bound state energies of quarkonia in the 
Born-Oppenheimer approximation.
We use input bottom- and charm-quark masses $m_b=4977\,\mathrm{MeV},
 m_c=1628\,\mathrm{MeV}$ from quark models 
\cite{Capitani:2018rox,Godfrey:1985xj}.
\begin{table}[h!]
\centering
\begin{tabular}{l | c c c c c}
\toprule
$m_Q$ & $l=0$ & $l=1$ & $l=2$ & $l=3$ & $l=4$ \\
\midrule
$m_b$ & 3 & 3 & 2 & 2 & 1 \\
$m_c$ & 2 & 1 & 1 & -- &  -- \\
\bottomrule
\end{tabular}
\caption{Number of bound-state solutions $E_{nl}<0$ for bottomonium and 
   charmonium of the Schr\"odinger equation. 
  \label{t:quarkonium}
}
\end{table}

The number of bottomonium bound states is listed in the first row of
Tab.~\ref{t:quarkonium}. We
observe three $S$-wave states in agreement with the physical spectrum of 
$\Upsilon$ mesons. We do not find bound states beyond $l>4$.
For the states closest to the threshold for each angular 
momentum, the size of the wave function as determined from the root
mean-square radius is between $0.63\,r_c$ and $0.79\,r_c$.
We also solved the Schr\"odinger equation
when setting $V(r)=\min\{\sigma r+\hat V_0,0\}$ and find changes in 
energies of less than $10\,\mathrm{MeV}$, within the errors from the fit
parameters, so the bound-state spectrum is not sensitive to the width of the
mixing region.
The second row of Tab.~\ref{t:quarkonium} shows the number of bound states
we find for charmonium for comparison.

This calculation is only a check that counting the bound states leads to
sensible results. Our model Eq.~(\ref{e:model}) provides input
for more refined calculations of the coupled-channel scattering of
heavy-light mesons and the decays of quarkonia around threshold.

\section{Conclusions}
\label{sec:con}

This work presents a first calculation of string breaking with
$N_\mathrm{f}=2+1$ dynamical quarks from lattice QCD on a single ensemble with
light/strange quark masses that are heavier/lighter than in nature, corresponding to
$m_{\pi}=280\,\mathrm{MeV}$ and $m_{K}=460\,\mathrm{MeV}$. 
We compute the three lowest energy levels of a static quark and
anti-quark pair and observe the avoided level crossings  due to pair-creation of light and strange sea quarks.
Our main result can be summarised in the model of Eq.~(\ref{e:model})
which provides a very good description of our spectrum, shown in 
Fig.~\ref{figure:model}. In particular, we extract the values of the parameters 
$g_1=47.2(1.4)\,\mathrm{MeV}$ and $g_2=24.5(1.6)\,\mathrm{MeV}$ which describe 
the mixing between the gluonic flux tube and the broken string.

Further analysis to study the dependence of string breaking
on the quark masses is in progress.

\section*{Acknowledgements}

We thank Tomasz Korzec for suggesting a test of the code of the correlation matrix.
The authors wish to acknowledge the DJEI/DES/SFI/HEA Irish Centre for High-End Computing (ICHEC) for the provision of computational facilities and support. We are grateful to our colleagues within the CLS initiative for sharing ensembles. The code for the calculations using the stochastic LapH method is built on the USQCD QDP++/Chroma library \cite{chroma}. BH was supported by Science Foundation Ireland under Grant No.~11/RFP/PHY3218. CJM acknowledges support by the U.S.~National Science Foundation under award PHY-1613449. VK has received funding from the European Union's Horizon 2020 research and innovation programme under the Marie Sklodowska-Curie grant agreement Number 642069.

\bibliographystyle{JHEP}
\bibliography{sb}

\providecommand{\href}[2]{#2}\begingroup\raggedright\begin{thebibliography}{10}

\bibitem{Wilson:1974sk}
K.~G. Wilson, {\it {Confinement of Quarks}},  {\em Phys. Rev.} {\bf D10} (1974)
  2445--2459. [,319(1974)].

\bibitem{Drummond:1998ar}
I.~T. Drummond, {\it {Strong coupling model for string breaking on the
  lattice}},  {\em Phys. Lett.} {\bf B434} (1998) 92--98,
  [\href{http://xxx.lanl.gov/abs/hep-lat/9805012}{{\tt hep-lat/9805012}}].

\bibitem{Philipsen:1998de}
O.~Philipsen and H.~Wittig, {\it {String breaking in nonAbelian gauge theories
  with fundamental matter fields}},  {\em Phys. Rev. Lett.} {\bf 81} (1998)
  4056--4059, [\href{http://xxx.lanl.gov/abs/hep-lat/9807020}{{\tt
  hep-lat/9807020}}]. [Erratum: Phys. Rev. Lett.83,2684(1999)].

\bibitem{Knechtli:1998gf}
{\bf ALPHA} Collaboration, F.~Knechtli and R.~Sommer, {\it {String breaking in
  SU(2) gauge theory with scalar matter fields}},  {\em Phys. Lett.} {\bf B440}
  (1998) 345--352, [\href{http://xxx.lanl.gov/abs/hep-lat/9807022}{{\tt
  hep-lat/9807022}}]. [Erratum: Phys. Lett.B454,399(1999)].

\bibitem{Knechtli:2000df}
{\bf ALPHA} Collaboration, F.~Knechtli and R.~Sommer, {\it {String breaking as
  a mixing phenomenon in the SU(2) Higgs model}},  {\em Nucl. Phys.} {\bf B590}
  (2000) 309--328, [\href{http://xxx.lanl.gov/abs/hep-lat/0005021}{{\tt
  hep-lat/0005021}}].

\bibitem{bali}
{\bf SESAM Collaboration} Collaboration, G.~S. Bali, H.~Neff, T.~Duessel,
  T.~Lippert, and K.~Schilling, {\it {Observation of string breaking in QCD}},
  {\em Phys.Rev.} {\bf D71} (2005) 114513,
  [\href{http://xxx.lanl.gov/abs/hep-lat/0505012}{{\tt hep-lat/0505012}}].

\bibitem{Bresenham}
J.~E. Bresenham, {\it Algorithm for computer control of a digital plotter},
  {\em IBM Systems Journal} {\bf 4} (March, 1965) 25--30.

\bibitem{Bolder:2000un}
B.~Bolder, T.~Struckmann, G.~S. Bali, N.~Eicker, T.~Lippert, B.~Orth,
  K.~Schilling, and P.~Ueberholz, {\it {A High precision study of the Q anti-Q
  potential from Wilson loops in the regime of string breaking}},  {\em Phys.
  Rev.} {\bf D63} (2001) 074504,
  [\href{http://xxx.lanl.gov/abs/hep-lat/0005018}{{\tt hep-lat/0005018}}].

\bibitem{morningstar}
C.~Morningstar, J.~Bulava, J.~Foley, K.~J. Juge, D.~Lenkner, et~al., {\it
  {Improved stochastic estimation of quark propagation with Laplacian Heaviside
  smearing in lattice QCD}},  {\em Phys.Rev.} {\bf D83} (2011) 114505,
  [\href{http://xxx.lanl.gov/abs/1104.3870}{{\tt arXiv:1104.3870}}].

\bibitem{donnellan}
M.~Donnellan, F.~Knechtli, B.~Leder, and R.~Sommer, {\it {Determination of the
  Static Potential with Dynamical Fermions}},  {\em Nucl.Phys.} {\bf B849}
  (2011) 45--63, [\href{http://xxx.lanl.gov/abs/[hep-lat/1012.3037]}{{\tt
  [hep-lat/1012.3037]}}].

\bibitem{hyp2}
A.~Hasenfratz and F.~Knechtli, {\it {Flavor symmetry and the static potential
  with hypercubic blocking}},  {\em Phys.Rev.} {\bf D64} (2001) 034504,
  [\href{http://xxx.lanl.gov/abs/hep-lat/0103029}{{\tt hep-lat/0103029}}].

\bibitem{dellamorte}
M.~{Della Morte}, A.~Shindler, and R.~Sommer, {\it {On lattice actions for
  static quarks}},  {\em JHEP} {\bf 0508} (2005) 051,
  [\href{http://xxx.lanl.gov/abs/hep-lat/0506008}{{\tt hep-lat/0506008}}].

\bibitem{eichtenhill}
E.~Eichten and B.~R. Hill, {\it {An Effective Field Theory for the Calculation
  of Matrix Elements Involving Heavy Quarks}},  {\em Phys.Lett.} {\bf B234}
  (1990) 511.

\bibitem{distillation}
{\bf Hadron Spectrum Collaboration} Collaboration, M.~Peardon et~al., {\it {A
  Novel quark-field creation operator construction for hadronic physics in
  lattice QCD}},  {\em Phys.Rev.} {\bf D80} (2009) 054506,
  [\href{http://xxx.lanl.gov/abs/0905.2160}{{\tt arXiv:0905.2160}}].

\bibitem{stout}
C.~Morningstar and M.~J. Peardon, {\it {Analytic smearing of SU(3) link
  variables in lattice QCD}},  {\em Phys.Rev.} {\bf D69} (2004) 054501,
  [\href{http://xxx.lanl.gov/abs/hep-lat/0311018}{{\tt hep-lat/0311018}}].

\bibitem{dilution2}
W.~Wilcox, {\it {Noise methods for flavor singlet quantities}},  in {\em
  {Numerical challenges in lattice quantum chromodynamics. Proceedings, Joint
  Interdisciplinary Workshop, Wuppertal, Germany, August 22-24, 1999}},
  pp.~127--141, 1999.
\newblock \href{http://xxx.lanl.gov/abs/hep-lat/9911013}{{\tt
  hep-lat/9911013}}.

\bibitem{dilution}
J.~Foley, K.~{Jimmy Juge}, A.~O'Cais, M.~Peardon, S.~M. Ryan, et~al., {\it
  {Practical all-to-all propagators for lattice QCD}},  {\em
  Comput.Phys.Commun.} {\bf 172} (2005) 145--162,
  [\href{http://xxx.lanl.gov/abs/hep-lat/0505023}{{\tt hep-lat/0505023}}].

\bibitem{andersen}
C.~Andersen, J.~Bulava, B.~Hörz, and C.~Morningstar, {\it {The $I=1$ pion-pion
  scattering amplitude and timelike pion form factor from $N_{\rm f} = 2+1$
  lattice QCD}},  {\em Nucl. Phys.} {\bf B939} (2019) 145--173,
  [\href{http://xxx.lanl.gov/abs/1808.05007}{{\tt arXiv:1808.05007}}].

\bibitem{gevp}
B.~Blossier, M.~{Della Morte}, G.~von Hippel, T.~Mendes, and R.~Sommer, {\it
  {On the generalized eigenvalue method for energies and matrix elements in
  lattice field theory}},  {\em JHEP} {\bf 0904} (2009) 094,
  [\href{http://xxx.lanl.gov/abs/0902.1265}{{\tt arXiv:0902.1265}}].

\bibitem{luscher_gevp}
M.~Lüscher and U.~Wolff, {\it How to calculate the elastic scattering matrix
  in two-dimensional quantum field theories by numerical simulation},  {\em
  Nuclear Physics B} {\bf 339} (1990), no.~1 222 -- 252.

\bibitem{bulava_phaseshifts}
J.~Bulava, B.~Fahy, B.~Hörz, K.~J. Juge, C.~Morningstar, and C.~H. Wong, {\it
  {$I=1$ and $I=2$ $\pi-\pi$ scattering phase shifts from $N_{\mathrm{f}} =
  2+1$ lattice QCD}},  {\em Nucl. Phys.} {\bf B910} (2016) 842--867,
  [\href{http://xxx.lanl.gov/abs/1604.05593}{{\tt arXiv:1604.05593}}].

\bibitem{cls}
M.~Bruno et~al., {\it {Simulation of QCD with N$_{f} =$ 2 $+$ 1 flavors of
  non-perturbatively improved Wilson fermions}},  {\em JHEP} {\bf 02} (2015)
  043, [\href{http://xxx.lanl.gov/abs/1411.3982}{{\tt arXiv:1411.3982}}].

\bibitem{openboundary}
M.~Lüscher and S.~Schaefer, {\it {Lattice QCD with open boundary conditions
  and twisted-mass reweighting}},  {\em Comput. Phys. Commun.} {\bf 184} (2013)
  519--528, [\href{http://xxx.lanl.gov/abs/1206.2809}{{\tt arXiv:1206.2809}}].

\bibitem{bootstrap1}
B.~Efron, {\it Bootstrap methods: Another look at the jackknife},  {\em The
  Annals of Statistics} {\bf 7} (1979), no.~1 1--26.

\bibitem{bootstrap2}
B.~Efron and R.~Tibshirani, {\em Bootstrap Methods for Standard Errors,
  Confidence Intervals, and Other Measures of Statistical Accuracy}, vol.~1.
\newblock Institute of Mathematical Statistics, 02, 1986.

\bibitem{Koch:2018puh}
V.~Koch, J.~Bulava, B.~{H\"{o}rz}, F.~Knechtli, G.~Moir, C.~Morningstar, and
  M.~Peardon, {\it {String breaking with 2+1 dynamical fermions using the
  stochastic LapH method}},  in {\em {36th International Symposium on Lattice
  Field Theory (Lattice 2018) East Lansing, MI, United States, July 22-28,
  2018}}, 2018.
\newblock \href{http://xxx.lanl.gov/abs/1811.09289}{{\tt arXiv:1811.09289}}.

\bibitem{bookfrancesco}
F.~Knechtli, M.~G{\"u}nther, and M.~Peardon, {\em {Lattice Quantum
  Chromodynamics: Practical Essentials}}.
\newblock SpringerBriefs in Physics. Springer, 2017.

\bibitem{cornell}
E.~Eichten, K.~Gottfried, T.~Kinoshita, J.~Kogut, K.~D. Lane, and T.~M. Yan,
  {\it Spectrum of charmed quark-antiquark bound states},  {\em Phys. Rev.
  Lett.} {\bf 34} (Feb, 1975) 369--372.

\bibitem{cls_scalesetting}
M.~Bruno, T.~Korzec, and S.~Schaefer, {\it {Setting the scale for the CLS $2 +
  1$ flavor ensembles}},  {\em Phys. Rev.} {\bf D95} (2017), no.~7 074504,
  [\href{http://xxx.lanl.gov/abs/1608.08900}{{\tt arXiv:1608.08900}}].

\bibitem{Capitani:2018rox}
S.~Capitani, O.~Philipsen, C.~Reisinger, C.~Riehl, and M.~Wagner, {\it
  {Precision computation of hybrid static potentials in SU(3) lattice gauge
  theory}},  {\em Phys. Rev.} {\bf D99} (2019), no.~3 034502,
  [\href{http://xxx.lanl.gov/abs/1811.11046}{{\tt arXiv:1811.11046}}].

\bibitem{Godfrey:1985xj}
S.~Godfrey and N.~Isgur, {\it {Mesons in a Relativized Quark Model with
  Chromodynamics}},  {\em Phys. Rev.} {\bf D32} (1985) 189--231.

\bibitem{chroma}
{\bf SciDAC, LHPC, UKQCD} Collaboration, R.~G. Edwards and B.~Joo, {\it {The
  Chroma software system for lattice QCD}},  {\em Nucl. Phys. Proc. Suppl.}
  {\bf 140} (2005) 832, [\href{http://xxx.lanl.gov/abs/hep-lat/0409003}{{\tt
  hep-lat/0409003}}]. [832(2004)].

\end{thebibliography}\endgroup

\end{document}